# Formation and Controlling of Optical Hopfions in High Harmonic Generation

Zijian Lyu[1], Yiqi Fang[2], and Yunquan Liu[1,3,4*]

[1]*State Key Laboratory for Mesoscopic Physics and Frontiers Science Center for Nano-Optoelectronics, School of Physics, Peking University, Beijing 100871, China*
[2]*Department of Physics, Universität Konstanz, Konstanz 78464, Germany*
[3]*Beijing Academy of Quantum Information Sciences, Beijing 100193, China*
[4]*Collaborative Innovation Center of Extreme Optics, Shanxi University, Taiyuan, Shanxi 030006, China*
(Dated: October 8, 2024)

Toroidal vortex is a kind of novel and exotic structured light with potential applications in photonic topology and quantum information. Here, we present the study on high harmonic generation (HHG) from the interaction of the intense toroidal vortices with atoms. We show that the spatial distribution of harmonic spectra reveal unique structures, which are kinds of high-order topological toroidal vortex because of the rotating of the driving fields with transverse orbital angular momentum. Then we show that, by synthesizing the toroidal vortices and optical vortices with longitudinal orbital angular momentum, it is able to generate the optical hopfions in extreme ultraviolet (EUV) range, which exhibit completely different topological property compared with both the driving fields. Harnessing the spin angular momentum selection rules, we further show that one can control the topological texture and channel each harmonic into a single mode of controllable Hopf invariant.

The study on topologically non-trivial defects, textures, and knots, etc., has inspired remarkably ubiquitous progress in physics, which can be dated back to 1860's when Lord Kelvin proposed his vortex atom hypothesis [1]. Ever since, the study of vortex ring has been proliferated in fluid dynamics [2, 3], magnetics [4] and photonics [5–7]. The recently discovered photonic toroidal vortex [7] is characterized by a torus-shaped structure in the space-time plane and possesses transverse orbital angular momentum (OAM) whose direction rotates around the torus ring. However, the closed vortex ring corresponds to the unknot which is topologically different from the linked knots in hopfions according to the knot theory. Very recently, the scalar optical hopfions in the shape of toroidal vortex has been demonstrated by weaving spatiotemporal optical vortex in linear optic regime[8].

Optical hopfions are kinds of typical structured light with three-dimensional (3D) topological states. Because of the unique field profiles and particle-like properties, they have been explored in many physical fields, including high-energy physics[9], quantum fields[10, 11], condensed matter physics[12–15] and cosmology[16, 17]. Recently, photonic hopfions have been realized in free space [8, 18, 19]. It was initially proposed in the Skyrme-Fadeev model [20–22], and consists of a set of torus knots which are disjoint and linked loops that nest together to form complete ring tori. This knot-like solitons can be elegantly mapped to Hopf fibration [23] and be characterized by two independent winding numbers [24, 25]. The Hopf fibration provides a map from the unit sphere in 4D space ($S^3$) to the unit sphere in 3D sphere ($S^2$), which belongs to the third homotopy group $\pi_3(S^2) = Z$. The preimages of each points in $S^2$ are disjoint and linked circles ($S^1$) in $S^3$. By using stereographic projection, it is possible to visualize $S^3$ in three-dimensional space while preserving the topological property of linkedness of closed loops [23].

High harmonic generation (HHG) is an extreme nonlinear optical up-conversion phenomenon mediated by the strong laser-matter interaction. For the atomic process, it is usually described with three-step recollision picture [26–28]. In frequency domain, harmonics of different order independently conserve the important physical quantities of photons, such as energy [29], linear momentum [30], and spin angular momentum [31], laying the foundation of wide applications of HHG. In these years, there have been exciting progresses in the field of HHG driven by the optical vortex with OAM [32–36]. There, the angular momentum conservation of photon longitudinal OAM and transverse OAM in HHG have been investigated [32–34, 37]. So far, almost all the HHG studies involving the structured lights are simply related with the up-conversion of the frequency of fundamental fields, without changing the topology as compared with the driving fields. The generation of arbitrary optical OAM with controllable topological texture for HHG, i.e., optical hopfions in extreme ultraviolet (EUV) regime, would be great interests in ultrafast physics community.

In this Letter, we study the controlling topological structure of light fields in extreme ultraviolet (EUV) range with HHG driven by the intense toroidal vortices. The conservation of photon orbital and spin angular momentum on arbitrary direction is discussed. We show that the HHG driven by toroidal vortices is tilt in spatial-resolved spectrum. In each order HHG, fine interference structure is introduced by the transverse OAM of the optical toroidal vortices. In time domain, the $n$th harmonic reveals the torus ring structure while its transverse OAM per photon $L_n$ is equal to $n$ times that of the fundamental toroidal vortices, i.e., $L_n = nl\hbar$. We further show that, when combining the toroidal vortex field with the vortex field with longitudinal OAM to drive the HHG, each order of the harmonics can astonishingly form the optical hopfions whose equaphase lines form closed and linked loops when constrained with countervorticity of spin angular momentum, which are topologically different from the unlinked loops in harmonics driven by the toroidal vortices. We find that the two winding numbers of hopfions correspond to the transverse and longitudinal orbital angular momentum, respectively. We envision that HHG with intense toroidal vortices will provide a very versatile toolbox for creating and controlling EUV radiation with novel topological structures.

The field distribution of the toroidal vortices can be expressed as [7]

$$E = E_0 A_0 \left(\frac{\sqrt{(r_\perp - r_0)^2 + \xi^2}}{w_s}\right)^l \exp\left[-\frac{(r_\perp - r_0)^2 + \xi^2}{w_s^2}\right]$$
$$\times \exp[i(-l\tan^{-1}(\frac{\xi}{r_\perp - r_0}) - \omega_0 t + k_0 z)] \quad (1)$$

where $r_\perp = \sqrt{x^2 + y^2}$, $r_0$ is the radius of the circular vortex line, $\xi = ct - z$ is a longitudinal coordinate local to the pulse, $c$ is the group velocity of light, $E_0$ is the peak electric field strength, $A_0$ is a normalization constant, $w_s$ is the waist of beam , and $\omega_0$ and $k_0$ are the center angular frequency and wave number, respectively. In Fig. 1(a), we show the iso-intensity surface of a photonic toroidal vortex of $l = 1$ at $z = 0$. In the calculation, we used an intense driving toroidal vortex at the wavelength of 400-nm wavelength ($\lambda_{400}$) at the peak intensity of $7 \times 10^{14} W/cm^2$, $w_s$ and $r_0$ are taken to be $3\lambda_{400}$ and $6\lambda_{400}$. The photonic toroidal vortex exhibits a three-dimensional phase structure that rotates along a closed loop, creating a ring-shaped vortex line. The red arrows inside the vortex indicate the direction of local angular momentum density. Fig. 1 (b) and (c) shows the spatiotemporal distribution of the electric field and spiral phase of the photonic toroidal vortex at the cross-section.

We simulate the process of high harmonic generation (HHG) in Argon atoms by the intense field of photonic toroidal vortice using strong-field approximation (SFA) [37, 38]. Considering the rotational symmetry of the vortex ring around its center, we show the calculated HHG at the cross-section $x = 0$ of toroidal vortex. The simulated spatial-resolved harmonics spectrum is shown in Fig.1(d). The harmonic spectra in the cross-section $x = 0$ plane have spatial chirp with opposite tilting directions in the upper and lower regions. This can be understood by considering the toroidal vortex as the spatiotemporal optical vortex (STOV) [39, 40] rotating around the center of the toroidal for one revolution. From this point of view, the upper and lower part of the toroidal vortex in the cross-section $x = 0$ plane possess opposite direction of transverse OAM. By conducting the Fourier transform of the HHG spectrum with respect to time, we can ob-tain the angular frequency spectrum of the HHG field at each position along the $r-$axis. For a toroidal vortex with $l = 1$, the analytical result is given by

$$F_\pm(\omega, r) \propto \exp[-\frac{(r-r_0)^2}{w_s^2}] \exp[-\frac{1}{4}(\frac{w_s}{c})^2(\omega - \omega_0)^2]$$
$$\times [\frac{2(r-r_0)}{w_s} \pm \frac{w_s}{c}(\omega - \omega_0)] \quad (2)$$

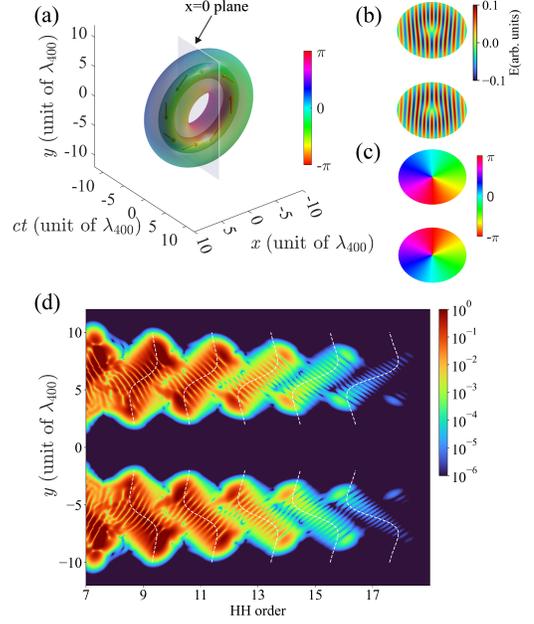

FIG. 1. (a) Side view of the 3D iso-intensity profile of a photonic toroidal vortex. The colors represent the spatiotemporal spiral phase ($\Phi_{r_\perp,t}$). The spiral phase has a topological charge of 1. The red arrows indicate the direction of local OAM density. The transparent rectangle denotes the location of the cross-section of plane $x = 0$. (b) spatiotemporal distribution of the strength of electric field $E(y,t)$ and (c) spiral phase of the photonic toroidal vortex $\varphi(y,t)$ at the cross-section plane of $x = 0$ ($y - t$ plane). (d) Spatial-resolved HHG driven by the electric field at the cross-section plane of of $x = 0$.

The '$\pm$' subscripts in $F_\pm$ corresponds the upper and lower parts of the harmonic spectrum, respectively. The spatiospectral weight of each angular frequency is determined by $F_\pm^2(r,\omega)$. By defining the local center angular momentum,

$$\omega_\pm(r) = [\int_{-\infty}^{\infty} F_\pm^2(r,\omega)\omega d\omega]/[\int_{-\infty}^{\infty} F_\pm^2(r,\omega)d\omega]$$
$$= \omega_0 + \Delta\omega_\pm(r) \quad (3)$$

The spatial chirp of the toroidal vortex can be quantitatively characterized through $\Delta\omega_\pm(r) = \pm\frac{\lambda_{400}}{2\pi}\frac{4(r-r_0)}{4(r-r_0)^2+w_s^2}\omega_0$, which takes the form of a dispersive lineshape. Due to the opposite transverse angular momentum directions of the upper and lower parts in the cross-section plane, the spectrum exhibits different tilting directions. This aspect resembles with the intensity spatial distribution of vortex beams with the longitudinal OAM focusing through a cylindrical lens[41]. The local center of angular frequency for $n$th harmonics $\omega_{n,\pm}$ can be connected with that of the fundamental frequency with $\omega_{n,\pm}(r) = n\omega_{\pm}(r) = n\omega_0 + n\Delta\omega_{\pm}(r)$. The $\omega_{n,\pm}(r)$ is plotted with the white dashed curves in Fig. 1(d), which agrees well with the spatiospectral tilt of the harmonics.

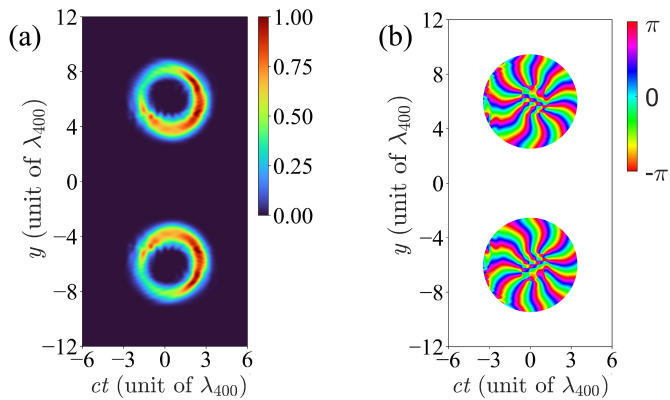

FIG. 2. The spatiotemporal profiles of the intensity (a) and phase (b) of the 11th harmonic at the cross-section ($y$-$t$ plane). The phase distribution indicates that the corresponding topological charge of the order harmonic is $l=11$.

Furthermore, on can see each part of EUV toroidal vortex exhibits distinct interference fringes in each harmonic. The fine interference patterns originate from the interference of the harmonics emitted before the phase singularity and after the singularity [37]. Besides, the number of fringes is equal to the topological charge $l$, which reflects the conservation of transverse OAM. To verify the general conservation law in this case, we illustrate the spatiotemporal intensity of 11th harmonic in Fig. 2(a) and its phase information in Fig. 2(b) from the calculated harmonic fields. The spatiotemporal profile of the of the 11th harmonic is shown in simulation [42]. We observe a well defined ring-shaped distribution for the intensity and a phase profile linearly increasing by $11 \times 2\pi$ along a circle centered on the phase singularity. The spatial asymmetry of intensity originates from the spatial chirp of the harmonics spectrum since as the radial $r$ increases, corresponding to higher harmonic frequencies, the harmonics yield decreases near the cutoff region. On the other hand, numerical calculations of the OAM per photon simulation [42], reveal that the transverse OAM values for the 11th and 13th harmonics are $10.92\hbar$ and $12.89\hbar$, respectively. The results indicate the conservation law of transverse OAM as $L = nl\hbar$. The generated harmonics of higher orders carry larger transverse OAM.

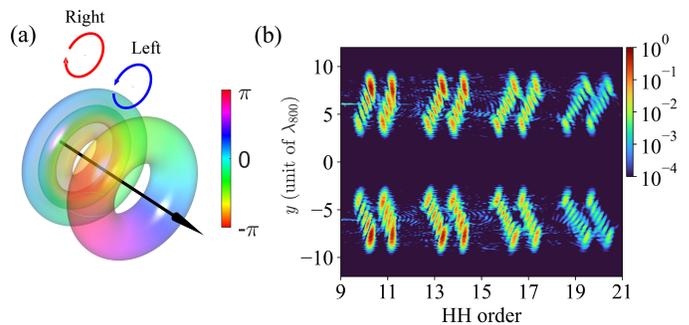

FIG. 3. (a) The iso-intensity profile of a toroidal vortex and a vortex. The radius of the toroidal vortex is set to $6\lambda_{800}$, and both the waist of the waist of the toroidal vortex and the vortex are set to $3\lambda_{800}$. The colors denote the spiral phase. The black arrow indicates the propagation direction. (b) high harmonic spectrum spectra in the cross section plane of $x = 0$ for the synthesied toroidal vortex and vortex with temporal overlapping. Here, $\omega'_0=0.057$ a.u. denotes the angular frequency of the 800-nm field. The peak electric field strength of 800- (toroidal vortex) and 400-(vortex) nm light fields in (a) and (b) are taken to be 0.12 a.u.

Up to now, the study on HHG with OAM has been limited to either pure longitudinal OAM $l_\parallel$ or pure transverse OAM $l_\perp$. In these cases, because of the parity and angular momentum conservation, the emitted EUV radiation can only embody the topological structure of driving lasers. To circumvent this limitation, we propose using a synthesized two-color ($\omega$ and $2\omega$) vortex fields to drive the HHG. As shown in Fig. 3(a), the fundamental field $\omega$ at 800 nm is a right-circular toriodal vortex $\sigma_\parallel = -1$ ($\sigma_\parallel$ being the spin of photons along the longitudinal direction) carrying transverse OAM $l_\perp = 1$ while the second harmonic of $2\omega$ at 400 nm is a left-circular vortex light with the longitudinal spin angular momentum $\sigma_\parallel = 1$ and OAM $l_\parallel = 1$ .

In Fig. 3(b), we present the simulated HHG spatial and energy spectra in the cross-section plane of $x = 0$. Since we introduce the constraints of spin angular momentum ($\sigma_\parallel$), the difference in the number of photons absorbed from each driving field must be one. This leads to the prohibition of the $3n$th harmonic, while the $(3n + 1)$th harmonic is contributed by the channel $n_{800} = n + 1, n_{400} = n$, and the $(3n - 1)$th harmonics is contributed by the channel $n_{800} = n - 1, n_{400} = n$ in which $n_{800}$ and $n_{400}$ are the numbers of photons absorbed from 800- and 400-nm fields [46]. Then, the longitudinal and transverse topological charges of $(3n+1)$th and $(3n - 1)$th harmonics are $(l_\perp = n + 1, l_\parallel = n)$ and $(l_\perp = n - 1, l_\parallel = n)$, respectively. One can find the number of fringes equals to the topological charge of $l_\perp$. More generally, assume the driven field possesses OAM of $(l_\perp = p, l_\parallel = 0)$ and $(l_\perp = 0, l_\parallel = q)$, the longitudi-nal and transverse topological charges of $(3n+1)$th and $(3n-1)$th harmonics will be $(l_\perp = p(n+1), l_\parallel = qn)$ and $(l_\perp = p(n-1), l_\parallel = qn)$, respectively.

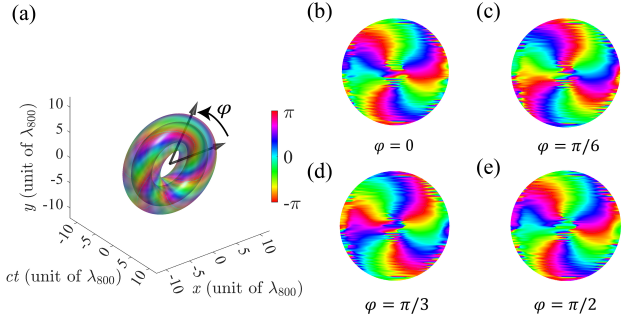
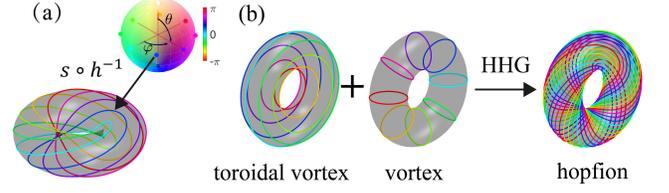

FIG. 4. Optical Hopfion of HHG: (a) spatiotemporal phase distribution on the iso-intensity surface of 11th harmonic. (b)-(e) circulating spiral phase in the radial-temporal plane (r-t plane).

FIG. 5. (a) In the parameter space, the longitude and latitude degrees ($\varphi$ and $\theta$) of a parametric 2-sphere ($S^2$) are represented by hue color and its lightness (the major radius and minor radius of the torus become smaller towards the South Pole and larger towards the North Pole.) The mapping $s \circ h^{-1}$ takes the points on the 2-sphere and maps them to closed equiphase lines on a torus surface, in which $h$ and $s$ correspond to Hopf map and stereographic projection, respectively. The lines projected from chosen points with the same latitude $\beta$ but varying longitude on the 2-sphere form a collection of tours knots that envelop a torus. (b) The equiphase lines of a toroidal vortex, conditional vortex and the 11th order harmonic. Different color represents different values of phase.

To further characterize the topological structure of the harmonics, we retrieve the spatiotemporal phase from the calculated 11th harmonic fields by inverse Fourier transform of the HHG spectrum [Fig. 3(b)]. Very interestingly, we find that the topological structure of each harmonic field reveals as optical hopfions. In Fig.4(a), we illustrates the phase distribution on the iso-intensity surface of 11th harmonic. It is made of the coupling between spatiotemporal spiral phase with topological charge $l_\perp = 3$ and spatial spiral phase with topological charge $l_\parallel = 4$ on iso-intensity surface. As shown in Fig.4(b)-(e), the spatial-temporal phase in the $r-t$ plane increases linearly by $3 \times 2\pi$ around the singularity. Furthermore, when the cross-section rotates along the vortex ring, the spatial-temporal phase also undergoes rotation with a period of $\pi/2$, indicating that the associated longitudinal OAM topological charge is equal to 4. In a more rigorous manner, we calculate $l_\perp$ and $l_\parallel$ of the winding numbers:

$$l_\perp = \frac{1}{2\pi} \oint_C \nabla \phi_{r,t}(r_{pol}) dr_{pol} \qquad (4)$$

$$l_\parallel = \frac{1}{2\pi} \oint_C \nabla \phi_{x,y}(r_{tor}) dr_{tor} \qquad (5)$$

where $r_{tor}$ and $r_{pol}$ are the position vectors in the toroidal plane and poloidal plane, respectively. The numerical calculations yield $l_\perp = 3$ and $l_\parallel = 4$. Thus, the generated harmonics correspond to optical hopfions with different winding numbers $(l_\perp, l_\parallel)$. Through HHG processes, high-order hopfions are generated. The adjacent harmonics with one order difference exhibit the same hopf invariant, defined as the product of the two winding numbers $Q_{Hopf} = l_\perp l_\parallel$. For instance, the 10th harmonic with winding number $(4,3)$ and 11th harmonic with $(3,4)$ both exhibit the same hopf invariant of 12. For higher order harmonics, the corresponding Hopf invariant will increase.

Another prominent feature of hopfions is that the equiphase lines are disjoint and linked closed loops on the toroidal surface. As shown in Fig.5(a), the hue color and lightness of each point on the parametric 2-sphere represents the latitude and longitude, $\phi$ and $\theta$, respectively. The preimages of each point on the parametric 2-sphere are disjoint and interlinked circles in units sphere in four-dimensional (4D) space ($S^3$). By employing stereographic projection, the $S^3$ structure that residing in 4D space can be perceptually revealed, preserving the topological property of linked closed loops. The loops formed by mapping points at the same latitude $\theta$ on the parametric sphere represents a set of torus knots, which cover the surface of a torus completely as points with different longitudes $\theta$ are scanned. The corresponding sterographic projection of an equiphase line in real space is shown with the same colors as in the parameter space. To further elucidate the unique topological properties of hopfions, we plot the equiphase lines on the iso-intensity surfaces of the synthesized fundamental toroidal vortex and the second harmonic vortex in Fig 5(b). One can clearly see that the loops are unlinked, indicating that they are topologically different from hopfions. Hopfions emerge through the HHG. Taking the 11th harmonic with winding number (3,4) for example, we present eight equiphase lines with different phase. The equiphase line is a knot wind-ing about the ring three times and about the hole four times. The closed loop painted in one specific color corresponds to a point in the parameter space. Each closed loop is an equiphase line and all equiphase loops form complete tori. The links and knots with a linking number are determined by the Hopf invariant of each order harmonic. The topological features of optical hopfions of HHG are controlled. Additionally, it is noteworthy to consider the propagation behavior of optical hopfions in free space in an actual experiment. As demonstrated in our simulation[42], the optical hopfions retain their unique topological structure after propagating over a certain distance($\sim 1$ m).

In conclusion, we have studied the HHG driven by intense toroidal vortices and the manipulation of arbitrary OAM of the harmonics. We show that the microscopic atomic electron dynamics is controlled concertedly, which is a robust scheme to generate topologically optical HHG hopfions. The photonic topology is transformed into high order hopfions in highly nonlinear optics regime. It is shown that the combination of transverse and longitudinal OAM gives rise to completely different topological properties. The interplay of spin state, transverse orbit state and longitudinal orbit state of photon is of paramount importance in physics. This work offers an intuitive physical picture of HHG process with toroidal vortex, paving the way for the generation and manipulation of EUV optical hopfions.The HHG optical hopfions hold the promising application for studying high-dimensional EUV topological states and for exciting and probing the genetic topological dynamics. It will also have implications in various other nonlinear optical processes such as, second harmonic generation [47] and four wave mixing process [48, 49] with the toroidal vortex field.

This work is supported by the State Key R&D Program (No. 2022YFA1604301) and National Natural Science Foundation of China (Nos.12334013, 92050201 and 92250306).